# Fourier transform based iterative method for x-ray differential phase-contrast computed tomography

Wenxiang Cong[1], A. Momose[2], and Ge Wang[1]
[1]School of Biomedical Engineering and Sciences, Virginia Tech, Blacksburg, VA 24061
[2]Department of Advanced Materials Science, The University of Tokyo

**Abstract:** Biological soft tissues encountered in clinical and pre-clinical imaging mainly consist of light element atoms, and their composition is nearly uniform with little density variation. Thus, x-ray attenuation imaging suffers from low image contrast resolution. By contrast, x-ray phase shift of soft tissues is about a thousand times greater than x-ray absorption over the diagnostic energy range, thereby a significantly higher sensitivity can be achieved in terms of phase shift. In this paper, we propose a novel Fourier transform based iterative method to perform x-ray tomographic imaging of the refractive index directly from differential phase shift data. This approach offers distinct advantages in cases of incomplete and noisy data than analytic reconstruction, and especially suitable for phase-contrast interior tomography by incorporating prior knowledge in a region of interest (ROI). Biological experiments demonstrate the merits of the proposed approach.

**1. Introduction:** Biological soft tissues encountered in clinical and pre-clinical imaging mainly consists of light element atoms, and their elemental composition is nearly uniform with little density variation. X-ray computed tomography (CT) has been widely used in the biomedical field but the contrast resolution has been poor along with the associated radiation dose problem [1]. Novel x-ray imaging research and development has been performed over the past years, especially in the area of x-ray phase contrast imaging, because the x-ray phase shift of soft tissues is about a thousand times greater than the x-ray attenuation coefficient over the diagnostic energy range. It is well recognized that phase-contrast imaging reveals detailed structural features of soft tissues and offers critical information that helps identify malignant from healthy tissues. Momose et al. first proposed two-grating interferometry in the hard X-ray region, and performed x-ray phase imaging using a phase transmission grating and an absorption transmission grating made by gold stripes on glass plates [2]. This work has been extended for 3D tomographic phase reconstruction using a hard x-ray two-grating interferometer [3-4]. Recently, three-grating interferometry in the hard x-ray region with low-brilliance x-ray tubes has been demonstrated [5]. It is exciting that phase contrast imaging can be efficiently performed with a conventional x-ray source, allowing widespread applications in biomedical imaging, industrial nondestructive testing, or security screening. Diffraction enhanced imaging (DEI) is another effective phase-contrast method in medical and biological fields. DEI takes advantage of high angular resolution of perfect crystals, has extremely high sensitivity with weakly absorbing low-Z samples, and produces refraction-angle images, which reflect gradients of the refractive index in the sample [6].

Differential phase-contrast tomography is to reconstruct a spatial distribution of the refractive index from differential projection data. The reconstruction problem of directional-derivative projection data has been studied for years. Pavlov et al. proposed that thephase shift could be recovered from differential data using a cubic spline approximation, and then the refractive index be reconstructed using the filtered-backprojection (FBP) method [7]. Maksimenko et al. proposed that the gradient of the refractive index could be first reconstructed, and then the refractive index be recovered using an integration method [8]. This type of integration methods was sensitive to noise and would generate strong streak noises [9]. Faris et al proposed a FBP method with a sign filter function directly applied to directional-derivative projection data for beam-deflection optical tomography [10]. This method was also adapted for phase-contrast tomography [11].

While analytic methods are widely used for image reconstruction, iterative reconstruction methods offer distinct advantages than analytic counterparts when data are incomplete, inconsistent, and rather noisy. Furthermore, because gratings of large sizes are difficult to fabricate and model in practice, x-ray grating interfermetric imaging is commonly performed wtih small samples [12]. Therefore, for large objects the interior tomography approach would be valuable to reconstruct a region of interest (ROI) accurately. Because the interior problem does not have a unique solution without any constraint [13], interior tomography achieves theoretically exact interior reconstruction by incorporating the prior aknowledge that are practically available.

In this paper, we propose a novel Fourier transform based iterative method to perform x-ray tomographic imaging directly from differential phase shift data. In the second Section, we will present the Fourier transform based iterative scheme, discuss the solution uniqueness, and describe an algorithm for interior differential phase-contrast tomography. In the third Section, experimatal results will be reported to demonstrate the feasibility and merits of the proposed approach. Finally, discussions and conclusion will be made.

**2. Materials and methods:** The experiments were performed using the beamline 20XU of SPring-8 at the synchrotron facility in Japan, where high-flux monochromatic x-rays have a sufficient spatial coherence. The x-ray Talbot interferometer was used that consists of a phase grating (G1), an amplitude grating (G2), and a charged-coupled device (CCD) with optical lens and a phosphor screen. The Talbot interferometer was arranged for the beamline and used in imaging biological soft tissues. In the Talbot interferometric imaging, moiré patterns were acquired using the fringe-scanning method. When one of the gratings was scanned along the transverse direction $x_g$, the intensity signal in each pixel in the detector plane would oscillate as a function of $x_g$. Then, the recorded individual intensity images at each projection view were processed for a differiantal phase shift image $\partial_s \Phi(s,\theta)$ [4].

**2.1. Iterative method:** The phase shift $\Phi(s,\theta)$ can be expressed as a projection of the refractive index along with the x-ray beam direction,

$$\Phi(s,\theta) = \frac{2\pi}{\lambda} \int_R \delta(s\theta + t\theta^\perp) dt, \qquad (1)$$

where $\lambda$ is the x-ray wavelength, $\theta$ a directional vector of the x-ray beam at a projection angle $\alpha$, and $\theta = (\cos\alpha, \sin\alpha)$. For x-ray grating interfermetric imaging, Eq. (1) is discretized into a linear system for each projection view,

$$\mathbf{A}_\theta \delta = \mathbf{B}_\theta \qquad (2)$$

where $\mathbf{B}_\theta$ is a vector of the discretized phase shift, $\mathbf{A}_\theta$ is weighting matrix, and $\delta$ is a vector of the discretized refractive index distribution to be reconstructed.

Then, we perform the discrete Fourier transform for both sides of Eq. (2), respectively,

$$\widetilde{\mathbf{A}}_\theta \delta = \widetilde{\mathbf{B}}_\theta \qquad (3)$$

where $\widetilde{\mathbf{A}}_\theta$ denotes the discrete Fourier transform of column vectors of $\mathbf{A}_\theta$, and $\widetilde{\mathbf{B}}_\theta$ is the discrete Fourier transform of $\mathbf{B}_\theta$. In term of the Fourier transform derivative theorem, we have

$$\partial_s \widetilde{\Phi}(s,\theta) = 2\pi i w\, \widetilde{\Phi}(w,\theta) \tag{4}$$

From Eqs. (3) and (4), we obtain the following linear system

$$\left(2\pi i w \widetilde{\mathbf{A}}_\theta\right)\delta = \widetilde{\mathbf{D}}_\theta \tag{5}$$

where $\widetilde{\mathbf{D}}_\theta$ is the discrete Fourier transform of differential phase shift data $\partial_s \Phi(s,\theta)$. With Eq. (5), we can reconstruct a refractive index distribution directly from differential phase shift data. Specifically, we can use a classical iterative method, such as the algebraic reconstruction technique (ART) or simultaneous algebraic reconstruction technique (SART).

**2.2. Interior reconstruction**: Most importantly, interior tomography can be performed based on Eq. (5) from truncated differential phase shift data. However, the interior problem is not uniquely solvable without appropriate prior knowledge. By incorporating a prior knowledge in a region of interest (ROI), the solution of the interior reconstruction of the refractive index is uniquely determined from truncated differential phase shift data, a mathematical discussion is presented as follows.

Without loss of generality, we assume that an object is supported on a disk $\Omega_{all} = \{x = (x_1, x_2) \in R^2 : |x| < A\}$, and an interior ROI defined as $\Omega_{ROI} = \{x = (x_1, x_2) \in R^2 : |x| < a\}$ for $0 < a < A$.

**Lemma 1** [14]: If $\delta(x)$ and $\delta_0(x)$ are compactly supported on the disk $\Omega_{all}$, and $\partial_s R\delta(s,\theta) = \partial_s R\delta_0(s,\theta)$, $-a \leq s \leq a$, $\theta \in S^1$, then the difference $u(x) = \delta(x) - \delta_0(x)$ is an analytic function in $\Omega_{ROI}$.

**Lemma 2** [15]: If a function $u(x)$ satisfies: $u(x) = 0$ and $Hu(x) = 0$ for $x \in (-a, a)$, where $Hu(x)$ is the Hilbert transform of $u(x)$, then $u(x) = 0$ for $x \in (-\infty, \infty)$.

From Lemmas 1 and 2, we can immediately obtain the following theorem on the solution uniqueness of interior refractive index reconstruction using the relation between the backprojection of differential projection of an image and its Hilbert transform [16]:

**Theorem 1.** If a refractive index image $\delta(x)$ is known on a small subregion $\Omega_{prior} \subset \Omega_{ROI}$ of ROI, then the refractive index function can be determined uniquely and stably from the truncated differential phase shift data $\partial_s \Phi(s,\theta)$ and the prior knowledge on $\Omega_{prior}$.

From Theorem 1, the interior reconstruction can be practically performed with an excellent image quality from truncated differential phase shift data and a known subregion inside an ROI based on Eq. (5) using an iterative method, such as ART and SART. Note that the prior knowledge must be incorporated into the iterative process so that the iteration would converge to the true refractive index distribution.

**3. Results and discussion:** To verify our method, we conducted a differential phase contrast tomographic imaging for a piece of rabbit liver. The biological sample was imaged using an x-ray grating Talbot interferometer at SPring-8 [4]. Using 0.1 nm x-rays, the acquisition of moiré patterns was done in a five-step fringe scan at each angular view of the sample rotation with a step of $0.72^0$ over a $180^0$ range Images were recorded on a CCD camera with optical lens and a phosphor screen. The CCD camera consisted of 706×706 pixels with an effective pixel size of 4.34 μm. From the recorded individual intensity images at every projection view, the differiantal phase shift images were extracted for each angular position of the sample.

**3.1. Global reconstruction:** To test the reconstruction robustness against measurement noise, we did not preprocess differiantal phase shift data for noise reduction. Representative reconstruction methods were implemented, including FBP directly from differiantal phase shift data, integration and reconstruction (first phase shift images were recovered from differiantal phase shift data using an integration method, and then the refractive index distribution reconstruction using either FBP or iteration methods), as well as our proposed Fourier transform based iterative method. In this comparative study, the image reconstruction using FBP appears noisy. The integration and reconstruction methods gave even worse results because the integration step generated strong artifacts, as shown in Fig. 1 (b). With the same differential projection dataset, 10 iterations were performed with the proposed method. We obtained an excellent refractive index image with highly contrast and spatial resolution, as shown in Fig.1 (c). This also shows that the proposed iterative method seems robust against measurement noise.

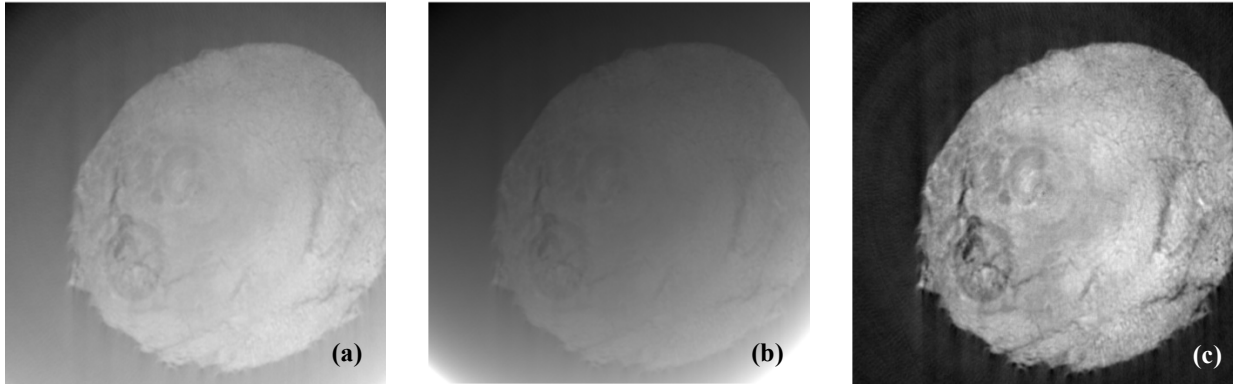

**Fig. 1.** Global reconstruction. (a) The image at the 200$^{th}$ slice reconstructed using FBP from differiantal phase shift data compromised by measurement noise; (b) the image reconstructed using the integration and reconstruction method; (c) the image reconstructed using the proposed Fourier transform based iterative method from the same dataset as that used in (a). The display window is [0.0, 1.1×10$^{-7}$].

**3.2. Interior reconstruction:** Furthermore, we conducted the interior reconstruction from truncated local differiantal phase shift data using the proposed iterative method. An ROI of the biological sample was selected to contain 128×128 pixels, which occupies only 3.3% of the global area. Nine pixel values around the ROI center were assumed as prior knowledge for interior reconstruction. Based on Eq. (5), 15 SART iterations were done incorporating the prior information. The reconstructed results are in an excellent agreement with the ROI in the global image reconstruction with FBP, as shown in Fig. 2. The detailed features in the ROI are quantitatively accurate.

**Conclusions:** In summary, we have proposed a Fourier transform based iterative method to perform the tomographic imaging in terms of the x-ray refractive index directly from differential phase shift data. Using experimental data from an x-ray grating interferometer in Japan, we have verified that the proposed method can accurately reconstruct an x-ray refractive index distribution and is robust against measurement noise. Specially, by incorporating the prior knowledge, this method is accurate and stable for interior reconstruction solely from truncated differential projection data. It is expected that this new method will help improve x-ray phase-contrast tomographic imaging and find biomedical applications.

**Acknowledgement:** The data used in this study was obtained in the experiment performed under the approval of the SPring-8 Committee 2005A0326-NM-np.

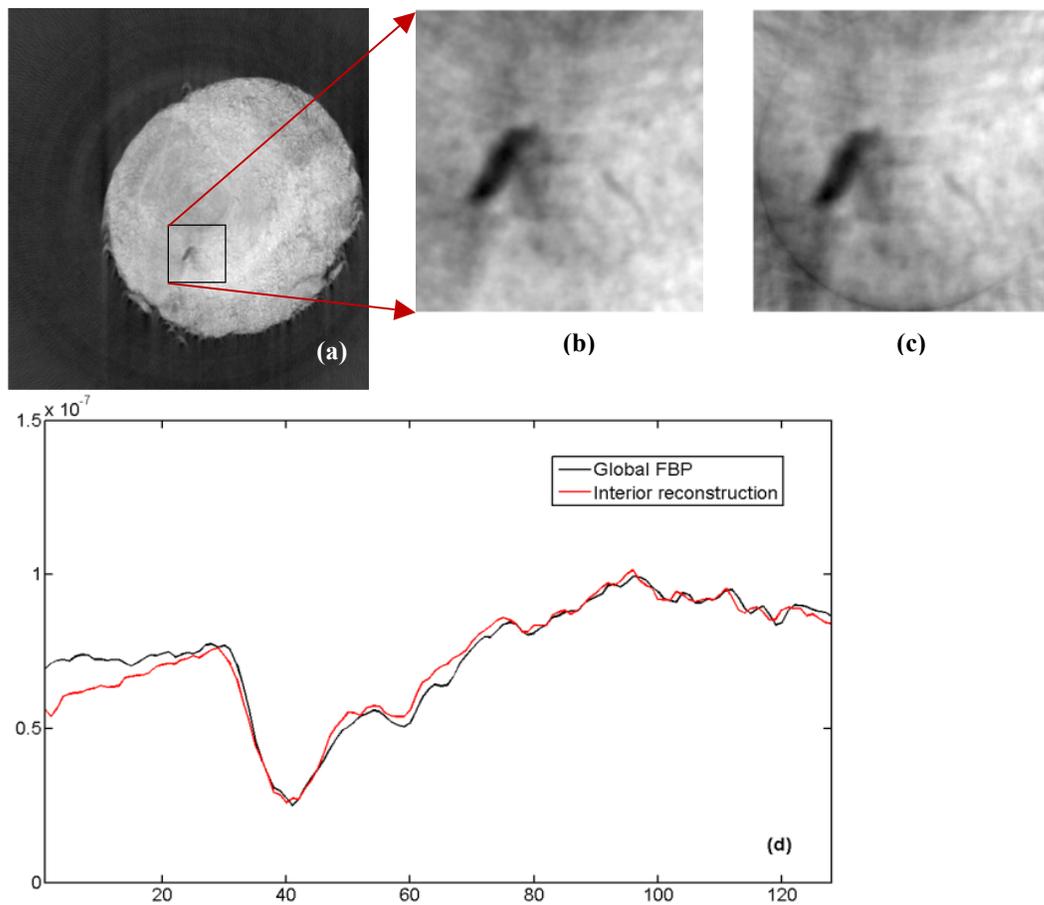

**Fig. 2.** Interior reconstruction. (a) The global image at the 250$^{th}$ slice reconstructed using FBP from differiantal phase shift data after the background noise is suppressed; (b) the magnified ROI in (a); (c) the interior image reconstructed at the same slice location using the Fourier transform based iterative method; (d) the profile comparison between (b) and (c) along the horizontal midline. The display window is [0.0, 1.1×10$^{-7}$].